# Spacer-layer-tunable magnetism and high-field topological Hall effect in topological insulator heterostructures

*Xiong Yao, *,† Hee Taek Yi, † Deepti Jain,§ Myung-Geun Han,⊥ and Seongshik Oh*,†*

[†]Center for Quantum Materials Synthesis and Department of Physics & Astronomy, Rutgers, The State University of New Jersey, Piscataway, New Jersey 08854, United States

[§]Department of Physics & Astronomy, Rutgers, The State University of New Jersey, Piscataway, New Jersey 08854, United States

[⊥]Condensed Matter Physics and Materials Science, Brookhaven National Lab, Upton, New York 11973, United States

*Email: xiong.yao@rutgers.edu

*Email: ohsean@physics.rutgers.edu

Phone: +1 (848) 445-8754 (S.O.)

ABSTRACT

Controlling magnetic order in magnetic topological insulators (MTIs) is a key to developing spintronic applications with MTIs, and is commonly achieved by changing the magnetic doping concentration, which inevitably affects spin-orbit-coupling strength and the very topological properties. Here, we demonstrate tunable magnetic properties in topological heterostructures over a wide range, from a ferromagnetic phase with Curie temperature of around 100 K all the way to a paramagnetic phase, while keeping the overall chemical composition the same, by controlling the thickness of non-magnetic spacer layers between two atomically-thin magnetic layers. This work showcases that spacer-layer control is a powerful tool to manipulate magneto-topological functionalities in MTI heterostructures. Furthermore, the interaction between the MTI and the $Cr_2O_3$ buffer layers also led to robust topological Hall effect surviving up to a record-high 6 T of magnetic field, shedding light on the critical role of interfacial layers in thin film topological materials.

Magnetic topological insulators (MTIs), incorporating both topology and magnetism, has greatly expanded the landscape of magnetic quantum materials in the past decade.[1-9] Most commonly, the magnetic order in MTIs is controlled by chemical doping but it entails the undesirable side effect of affecting the spin-orbit coupling (SOC) strength as well.[10] One way to manipulate the magnetic order in TIs with minimal compositional change is relying on magnetic proximity coupling: however, this approach suffers from very limited tunability.[11-14] An alternative and potentially more effective way to tune the magnetic properties in MTIs with minimal disruption of the chemical composition is to utilize a spacer-layer engineering scheme as follows.

Non-magnetic spacer layers are commonly used to tune the magnetic properties in transition metal and rare-earth metal multilayers,[15-17] which led to the famous giant magnetoresistance.[18, 19] Non-magnetic spacer layers can also effectively manipulate the interlayer coupling strength in MTIs, as reported in recent studies on $MnBi_{2n}Te_{3n+1}$ bulk crystals.[20-22] In such bulk crystals, however, change in the spacer-layer thickness inevitably results in change in the overall chemical composition. Here, we provide the first systematic study of spacer-layer-tunable magnetism in MTI heterostructures while keeping the overall chemical composition intact, and show that the magnetic ordering can be tuned from a ferromagnetic (FM) phase with Curie temperature of around 100 K all the way to a paramagnetic (PM) phase.

In order to systematically control the magnetic properties in MTIs, we grew multilayer heterostructures as shown in Figure 1a by molecular beam epitaxy (MBE). Two $CrBi_{0.32}Sb_{0.68}Te_3$ (CrBST) layers, each with a quintuple-layer (QL) thickness acting as a magnetic unit, were separated by non-magnetic $(Bi_{0.32}Sb_{0.68})_2Te_3$ (BST) spacer layers. We kept the total thickness (t + d = 8 QL) of the non-magnetic BST layers fixed, and changed only the thickness (t QL) of the spacer layer. We used epitaxial $Cr_2O_3$ buffer and amorphous $Cr_2O_3$ (a-$Cr_2O_3$) capping layer as the bottom and top layer, respectively, to improve the sample quality and avoid the aging effect: details can be found in "**Growth Method**" section. Figure 1b shows the high-angle annular dark-field scanning transmission electron microscopy (HAADF-STEM) of a heterostructure sample, which exhibits sharp interfaces and well-defined quintuple-layer structures of the MTI layers. The temperature-dependent sheet resistance $R_{xx}$ of five samples with t varying from 0 to 8 are depicted in Figure 1c, exhibiting semiconducting behavior commonly observed in MTIs.[1-3]

Figure 2a-e give the Hall resistance of the heterostructure samples measured at different temperatures. At the fixed temperature 6.7 K, the coercive field $H_C$ decreases with increasing spacer thickness and vanishes at t = 6 and 8. The nonlinear, yet non-hysteretic Hall resistance curves of t = 6 and 8 samples suggest that these samples are (close to) paramagnetic without long-range FM order. Figure 2f shows coercive fields extracted from Figure 2a-e. As shown in Figure S2, the Curie temperatures $T_C$, determined by the point where the coercive field approaches zero, monotonically decreases with increasing spacer thickness from around 100 K at t = 0 all the way down to 0 K (paramagnetic) at or beyond t = 6.

Figure 3a,b give the Hall ($R_{xy}$) and longitudinal sheet resistance ($R_{xx}$) measured at 1.5 K and magnetic field up to 9 T. The systematic evolution of $R_{xy}$ is consistent with Figure 2: the coercive field gradually decreases with increasing t and eventually vanishes at t = 6. $R_{xx}$ also exhibits a similar trend, showing butterfly-shaped hysteretic magnetoresistance (MR) at low magnetic field, which is characteristic feature of FM materials, for t = 0, 2, and 4 and non-hysteretic MR for t = 6 and 8. Nonetheless, the weak localization behavior at low magnetic field survives up to t = 6 and transitions to weak antilocalization at t = 8.[23] This suggests that t = 6 sample may still have an FM order, albeit with very small domain sizes resulting in vanishing hysteresis loop, and only t = 8 sample is clearly in the PM state.

Evidently, increasing interlayer distance between the Cr ions should result in reduced interlayer exchange coupling, which should obviously suppress FM ordering. Nonetheless, it should also be noted that as 't' increases in Figure 1a, the CrBST layers move away from the middle of the bulk and get closer to the surfaces: depending on the mechanism behind the FM ordering, this second effect could either enhance or suppress the FM ordering, as we will discuss below. In order to probe how the position of the CrBST layers with respect to the surface layer affects the magnetism, we synthesized three t = 0 samples with the two CrBST layers at three different locations - bottom, center and top - of the TI layers and measured their Hall effect in Figure S5. It shows that the FM order does depend significantly on the position of the CrBST layers, getting weaker as they move from the middle of the TI layers toward the surfaces.

In order to explain these observations, we consider two widely studied competing mechanisms: Ruderman–Kittel–Kasuya–Yosida (RKKY) vs Vleck mechanism.[10, 24-28] Under the RKKY mechanism, the exchange coupling between local magnetic ions is mediated by conduction electrons, so the strength of FM order strongly depends on the carrier density. On the other hand, under the van Vleck mechanism, the exchange coupling is mediated by bulk valence electrons, whose magnetic susceptibility is enhanced by the strong SOC strength of the bulk, so the strength of FM order does not depend on the carrier densities. A good example highlighting the role of bulk van Vleck mechanism behind FM ordering in a TI system is the FM-to-PM transition in Cr-doped $Bi_2(Se_xTe_{1-x})_3$ thin films induced by bulk band topology change as a function of x.[10]

Considering that RKKY is carrier-mediated while van Vleck is carrier-independent, a standard way to distinguish between the two is to probe how the FM order changes with the carrier density. As electric gating experiment is not applicable to our samples, we exploit aging effect to probe the effect of the carrier density on the FM order. Figure S3 and S6 depict the Hall resistance of t = 0 and 2 samples measured over long periods of time. Even though the carrier densities have clearly changed over time for both samples, the coercive field and Curie temperature remain almost unchanged. The carrier-independent FM ordering mechanism is also supported by the very observation that the strength of FM order as characterized by $H_C$ and $T_C$ is monotonically changing from t = 0 through 8 in Figure 2, even though their carrier densities and types are non-monotonically varying among these samples due to uncontrollable variations in the interfacial defect profiles. All these observations consistently point toward the van Vleck rather than the RKKY as the main mechanism for the observed FM order. Within this van Vleck picture, the weakening FM order as the CrBST layers move away from the middle of TI layers toward the surfaces in Figure S5 can be understood as a result of the diminishing bulk contribution to the FM order, which is the core of the bulk van Vleck mechanism. Accordingly, the wide tunability of the FM ordering strength with the spacer-layer control is likely due to the combined effect of the SOC-enhanced bulk van Vleck mechanism and the interlayer coupling, both of which go hand in hand as the spacer layer thickness changes. If it were of the RKKY type, the proximity to the conducting surface states should enhance intralayer coupling, thus compensating for the reduced interlayer coupling as the CrBST layers approach the surfaces: then, the magnetic tunability should be much suppressed or the FM ordering trend could even be reversed depending on whichever (interlayer vs intralayer coupling) wins.

In addition to the magnetic phase transition, another notable feature in Figure 3a is the hump-like feature in the Hall resistance of samples t = 0, 2, 4 and 6. We ascribe this nonlinearity to the topological Hall effect (THE) as explained below. Generally, the Hall resistance of a magnetic material is the sum of three contributions: the ordinary Hall resistance $R_{OH}$ induced by the deflection of carriers under magnetic field, the anomalous Hall resistance $R_{AH}$ induced by magnetization and momentum-space Berry curvature, and the topological Hall resistance $R_{TH}$ arising from real-space Berry curvature, as expressed below:

$$R_{xy} = R_{OH} + R_{AH} + R_{TH} \tag{1}$$

The $R_{OH}$ can be estimated by a linear extrapolation of the high field Hall resistance as $R_{AH}$ and $R_{TH}$ are significantly suppressed under high field. After subtracting $R_{OH}$, the resultant $R_{xy}$-$R_{OH}$ data presented in Figure 4a show antisymmetric "hump" features, which have been widely interpreted as the signature of THE,[29-36] the Hall response directly related to the Berry curvature in real-space. To single out the exact value of THE, we use $\tanh(H/H_0)$ as a Weiss molecular field theory approximation of $R_{AH}$,[36, 37] as shown in Figure 4a. After subtracting $R_{AH}$, the topological Hall resistance $R_{TH}$ clearly shows broad humps in Figure 4b. THE arises from the effective magnetic field felt by electrons traveling through spin textures, which is driven by the Dzyaloshinskii–Moriya (DM) interaction that tends to whirl the spins. In particular, the paramagnetic t = 8 sample does not exhibit any signature of THE after subtracting ordinary Hall resistance, as shown in Figure S9. The absence of THE in PM phase means that FM order is necessary for the formation of spin textures in this system.[35] Another notable feature as shown in Figure S10 is that the homogeneously Cr-doped BST sample also shows clear THE signal, implying that THE observed in our system is not related to the specific layering structure.

The appearance of a finite DM interaction requires both strong SOC strength and inversion-symmetry breaking. We would like to point out that our system shown in Figure 1a has broken inversion symmetry despite the presence of $Cr_2O_3$ layers on both top and bottom surfaces, because the bottom $Cr_2O_3$ buffer layer is epitaxial (having AFM order) while the top $Cr_2O_3$ capping layer is amorphous (thus, no AFM order). To clarify this point, we grew a series of control samples as shown in Figure S7. The composition of BST/CrBST layers in these sample are kept all the same. We can clearly see that the THE feature only appears in samples with epitaxial $Cr_2O_3$ buffer layer but is not related to the amorphous $Cr_2O_3$ capping layer. In particular, the stark contrast between Figure S7b (having only $Cr_2O_3$ buffer layer) and c (having only a-$Cr_2O_3$ capping layer) unambiguously shows that the interaction with $Cr_2O_3$ buffer and capping is quite different.

When compared to previously reported THE in MTIs,[29, 30, 34, 35] THE observed in our system shows couple of unique features. THE observed in Cr-doped BST or Mn-doped $Bi_2Te_3$ grown on non-magnetic substrates saturated at a field much lower than 1 T [29, 30, 35] and even in hard ferromagnetic V-doped $Sb_2Te_3$, THE did not survive beyond 2 T.[34] In contrast, the THE features in our system survive to much higher field of around 6 T even though the coercive field of our system is similar to those of typical Cr-doped BST system as shown in Figure 4. In high magnetic field range, the strength of THE is determined by the competition between magnetic field induced Zeeman energy and DM interaction.[34] As indicated by the control samples shown in Figure S7, the much enhanced THE in our system is likely due to the strong DM interaction fostered by the AFM $Cr_2O_3$ interfacial layers. According to a phenomenological model[34] (Figure S8), the effective DM interaction in our system is found to be of ~150 μeV, which is significantly larger than previous values.[34] Absence of any proximity-induced anomalous Hall effect (Figure S1a) suggests that the interfacial layer of the AFM $Cr_2O_3$ layer should be composed of random magnetic domains. Considering that the magnetic domain boundaries provide the real-space Berry curvature via DM interaction, the high-field THE observed in our system is likely due to the AFM domain boundaries at the $Cr_2O_3$ interfacial layers. Although further studies will be needed to fully understand the origin of this very high field THE, our observation paves a way toward engineering THE on MTIs via magnetic interfacial layers and substrates.

In summary, we demonstrated a wide range of magnetic tunability utilizing a spacer-layer control on MTI heterostructures. Using the nonmagnetic TI spacers as a "knob", we continuously tuned the magnetic order from an FM state with $T_C$ around 100 K all the way to a PM state without changing the chemical composition. The unparalleled magnetic tunability in this system unveils this material engineering scheme is a powerful tool to creating new magnetic states in thin film topological materials.[4, 5, 38] Moreover, interaction between the MTI and the AFM $Cr_2O_3$ buffer layers significantly enhanced the DM interaction and led to THE up to a record-high magnetic field of 6 T, providing a unique platform for high-field topological spintronic applications.

**Growth method**: All samples were grown on 10 mm × 10 mm $Al_2O_3$ (0001) substrates using a custom built SVTA MOS-V-2 MBE system with base pressure of low $10^{-10}$ Torr. 99.999% pure elemental Bi, Sb, Te and Cr sources were thermally evaporated using Knudsen cells for film growth. All the source fluxes were calibrated in situ by quartz crystal micro-balance (QCM) and ex situ by Rutherford backscattering spectroscopy (RBS). Substrates were cleaned ex situ by 5 minutes exposure to UV-generated ozone and in situ by heating to 750 ºC under oxygen pressure of $1 \times 10^{-6}$ Torr for ten minutes. Then the substrate was cooled down to 700 ºC and $Cr_2O_3$ buffer layer was grown by supplying Cr flux in $2 \times 10^{-6}$ Torr of oxygen. The sample was then cooled down to 260 ºC for the deposition of BST and CrBST layers. The ratio of Te flux to (Cr,Bi,Sb) flux were maintained at 10:1. The a-$Cr_2O_3$ capping layer was grown after the sample was cooled down to room temperature.

**Transport measurement:** All transport measurements were performed using the standard van der Pauw geometry, by manually pressing four indium wires on the corners of each sample. Two cryogenic systems were used for the transport measurements, including a closed-cycle cryostat with base temperature of 6.7 K and magnetic field up to 0.6 T, and a liquid He cryogenic system (AMI) with base temperature of 1.5 K and magnetic field up to 9 T. Raw data of $R_{xx}$ and $R_{xy}$ were properly symmetrized and anti-symmetrized, respectively.

**STEM measurement:** The cross-sectional TEM sample was prepared by focused ion beam with 5 keV Ga+ ion for final cleaning. STEM-HAADF images were obtained using an aberration-corrected JEOL ARM 200CF microscope at Brookhaven National Laboratory. The range of collection angle was 68 ~ 280 mrad.

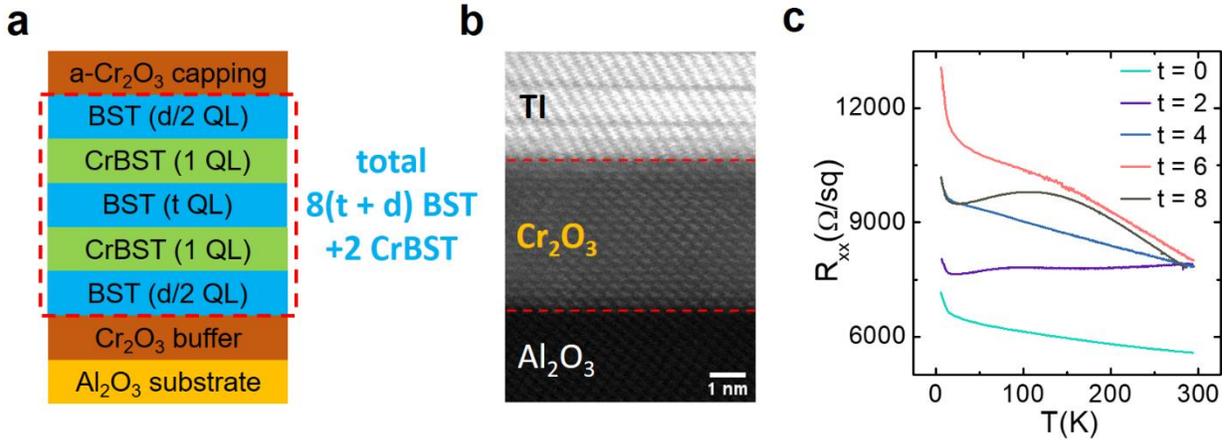

Figure 1: Structure and longitudinal sheet resistance of the CrBST/BST heterostructures. (a) Schematic of the CrBST/BST heterostructures. The spacer BST layers are t QL and the outer BST layers are d QL thick with the total BST layers being 8 QL thick. (b) High-angle annular dark-field scanning transmission electron microscopy (HAADF-STEM) image of a heterostructure sample. Red dash lines indicate the atomically sharp interfaces. (c) Temperature-dependent longitudinal sheet resistance measured down to 6.7 K for samples t = 0, t = 2, t = 4, t = 6, and t = 8.

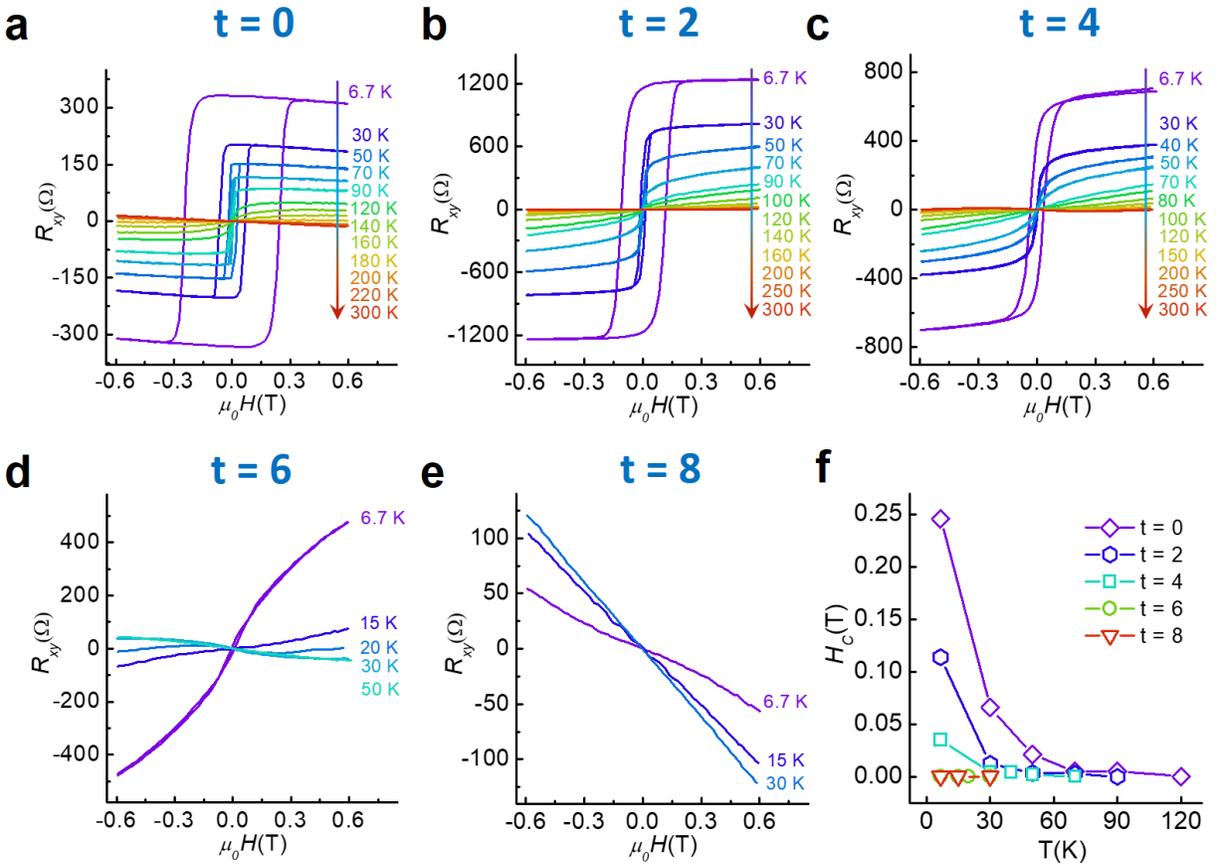

Figure 2: Hall resistance measured at different temperatures for the CrBST/BST heterostructures. (a-e) Field-dependent Hall resistance measured down to 6.7 K for samples t = 0 (a), t = 2 (b), t = 4 (c), t = 6 (d) and t = 8 (e). (f) Summarized coercive fields extracted from (a-e) for all the samples.

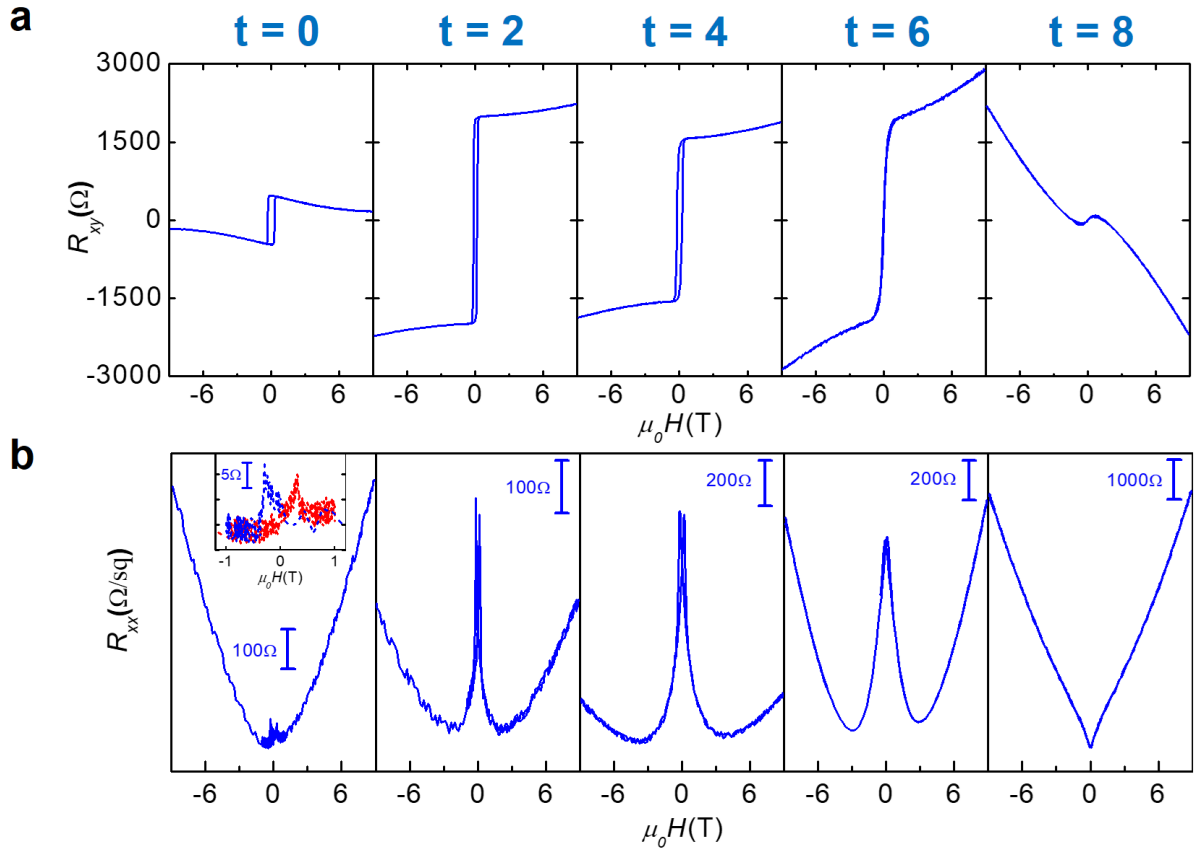

Figure 3: Transport results of the CrBST/BST heterostructures measured at 1.5 K and magnetic field up to 9 T. (a) Field-dependent Hall resistance measured at 1.5 K. (b) Field-dependent longitudinal sheet resistance measured at 1.5 K. Inset of sample t = 0 shows the enlarged plot at low field range.

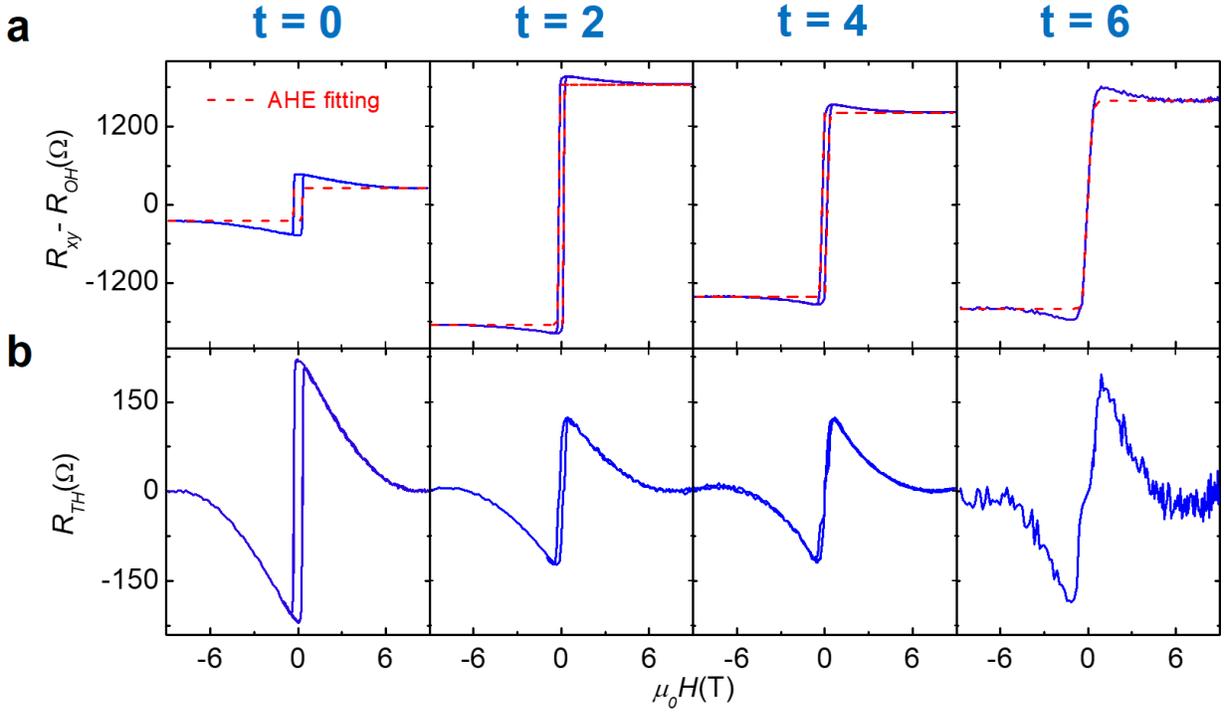

Figure 4: Topological Hall resistance of the CrBST/BST heterostructures measured at 1.5 K and magnetic field up to 9 T. (a) Hall resistance after subtracting ordinary Hall resistance for samples t = 0, t = 2, t = 4 and t = 6. The red dash lines show the fitting of anomalous Hall resistance by $\tanh(H/H_0)$. (b) Topological Hall resistance determined by subtracting both ordinary Hall resistance and anomalous Hall resistance for samples t = 0, t = 2, t = 4 and t = 6.

ASSOCIATED CONTENT

**Supporting Information**.

BST control samples with and without $Cr_2O_3$ buffer/capping layer, Curie temperatures for samples with different t, Hall resistance data of t = 0 and t = 2 samples as grown and months later, topological Hall resistance of sample t = 0 at 1.5 K and 33 K, control samples for testing the effect of CrBST layer position, comparison of Hall effect for t = 0 sample over time, control samples for testing the relationship between THE and $Cr_2O_3$ buffer/capping layer, fitting of the topological Hall effect, Hall resistance data for sample t = 8 before and after subtracting ordinary Hall resistance, transport data for a homogeneously Cr-doped BST sample. This material is available free of charge via the Internet at http://pubs.acs.org.

AUTHOR INFORMATION

Corresponding Authors

*E-mail: xiong.yao@rutgers.edu

*E-mail: ohsean@physics.rutgers.eduAuthor Contributions

X.Y. and S.O. conceived the experiments. X.Y., H.Y. and D.J. grew the thin films. X.Y. performed the transport measurements and analyzed the data with S.O. M.H. performed the STEM measurements. X.Y. and S.O. wrote the manuscript with contributions from all authors.

Notes

The authors declare no competing financial interest.

ACKNOWLEDGMENT

This work is supported by the center for Quantum Materials Synthesis (cQMS), funded by the Gordon and Betty Moore Foundation's EPiQS initiative through grant GBMF6402, and by Rutgers University. It is additionally supported by MURI W911NF2020166 and Army Research Office (ARO) Grant No. W911NF-20-1-0108. The work at Brookhaven National Laboratory is supported by the Materials Science and Engineering Divisions, Office of Basic Energy Sciences of the U.S. Department of Energy under contract no. DESC0012704. FIB use at the Center for Functional Nanomaterials, Brookhaven National Laboratory is acknowledged.

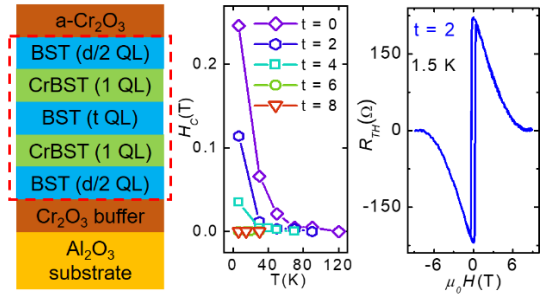

For TOC only

# *Supporting Information:*

# Spacer-layer-tunable magnetism and high-field topological Hall effect in topological insulator heterostructures

*Xiong Yao, [*,†] Hee Taek Yi,[†] Deepti Jain,[§] Myung-Geun Han,[⊥] and Seongshik Oh[*,†]*

[†]Center for Quantum Materials Synthesis and Department of Physics & Astronomy, Rutgers, The State University of New Jersey, Piscataway, New Jersey 08854, United States

[§]Department of Physics & Astronomy, Rutgers, The State University of New Jersey, Piscataway, New Jersey 08854, United States

[⊥]Condensed Matter Physics and Materials Science, Brookhaven National Lab, Upton, New York 11973, United States

*Email: xiong.yao@rutgers.edu

*Email: ohsean@physics.rutgers.edu

Phone: +1 (848) 445-8754 (S.O.)

## 1. BST control samples with and without $Cr_2O_3$ buffer/capping layer

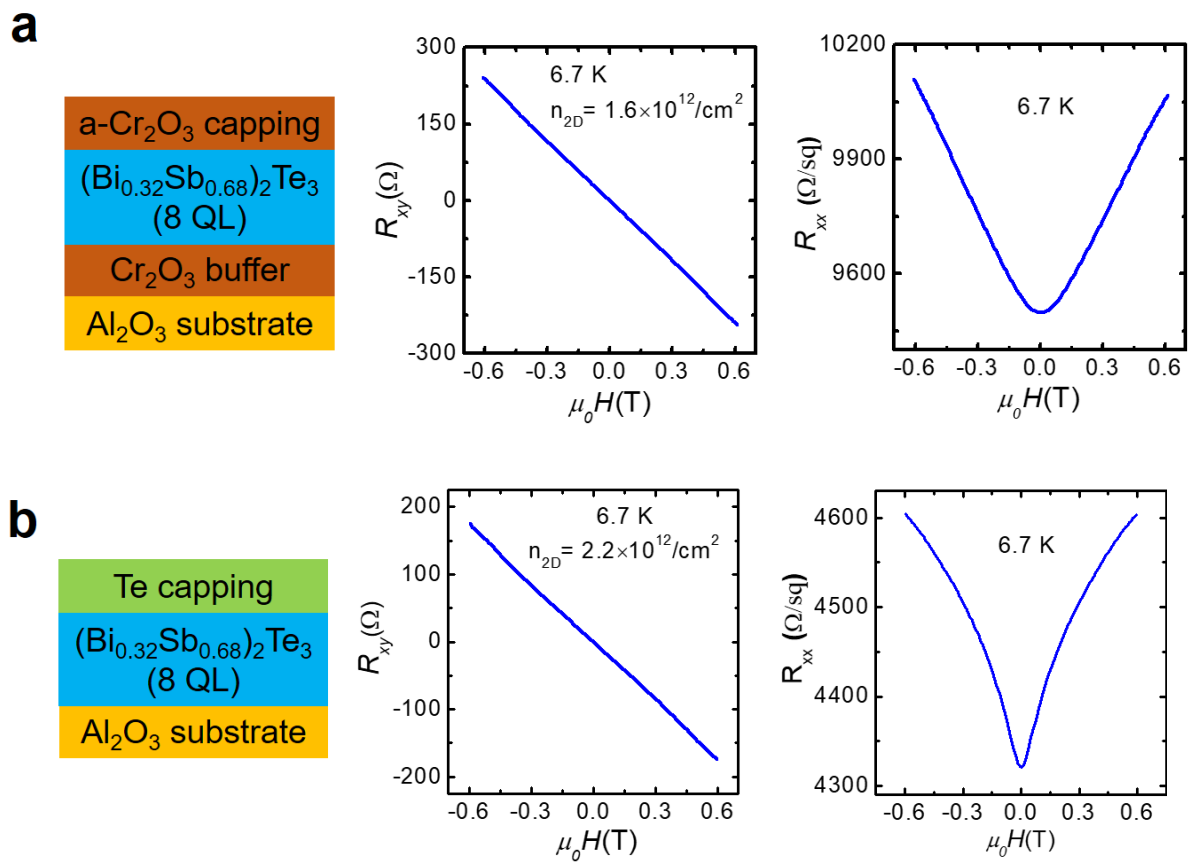

Figure S1: Schematic structure, field-dependent $R_{xy}$ and $R_{xx}$ for BST control samples (a) with and (b) without $Cr_2O_3$ buffer and capping layer.

## 2. Curie temperatures for samples with different t

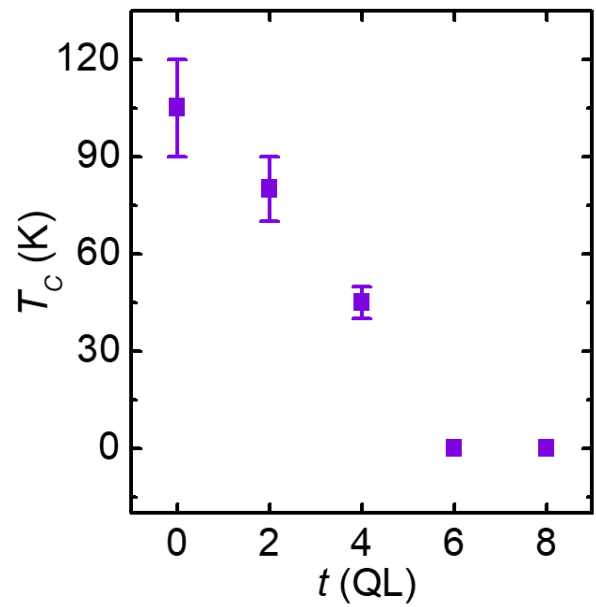

Figure S2: Curie temperatures determined by the point when coercive field approaches zero for samples with different t values.

## 3. Hall resistance data of t = 0 and t = 2 samples as grown and months later

**a**          t = 0

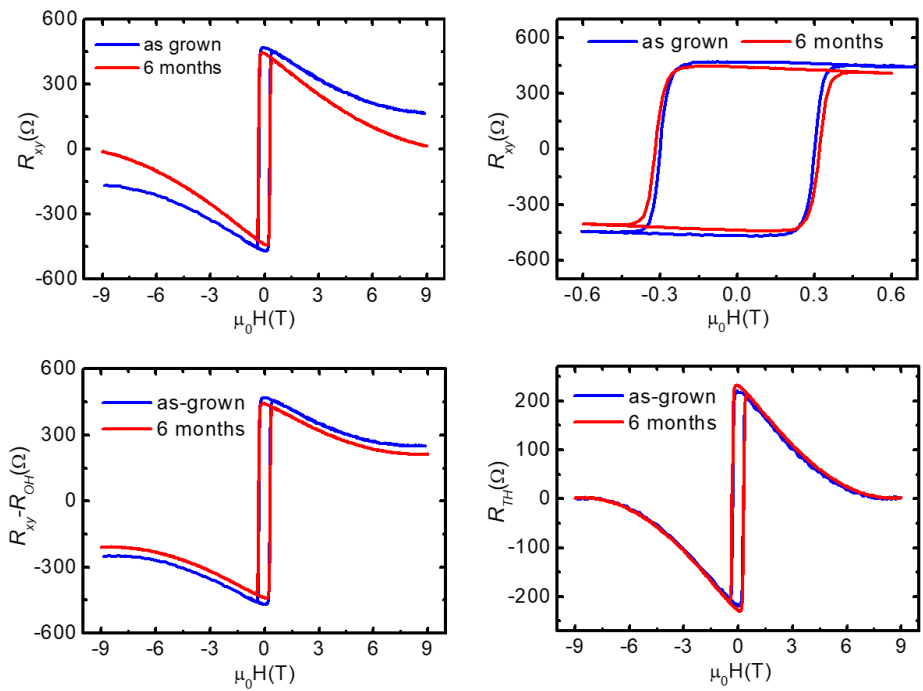

**b**          t = 2

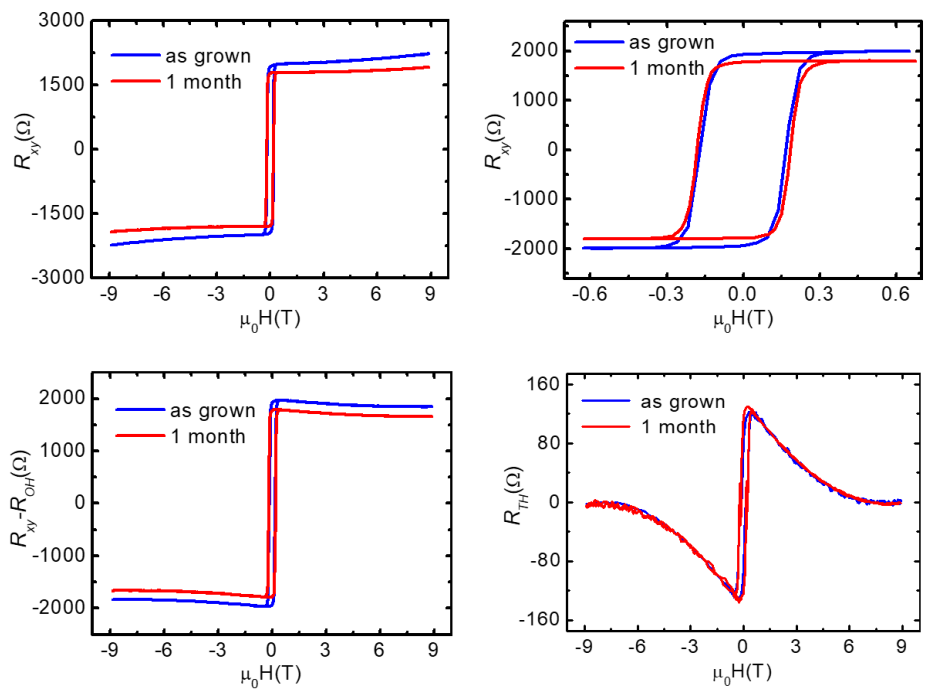

Figure S3: Field-dependent Hall resistance, after subtracting ordinary Hall resistance, after subtracting ordinary Hall resistance and anomalous Hall resistance measured at 1.5 K of (a) sample t = 0 as grown and 6 months later, (b) sample t = 2 as grown and 1 month later.

**4. Topological Hall resistance of sample t = 0 at 1.5 K and 33 K**

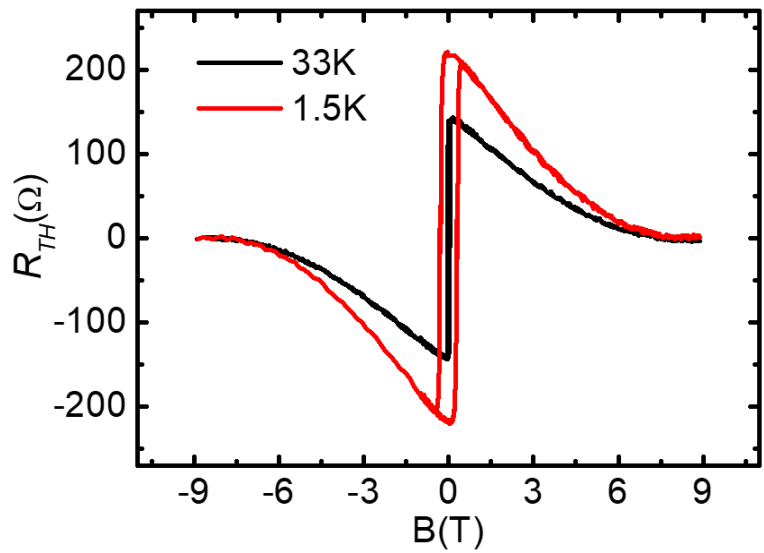

Figure S4: Topological Hall resistance after subtracting ordinary Hall resistance and anomalous Hall resistance of sample t = 0 at 1.5 K and 33 K respectively.

## 5. Control samples for testing the effect of CrBST-layer position

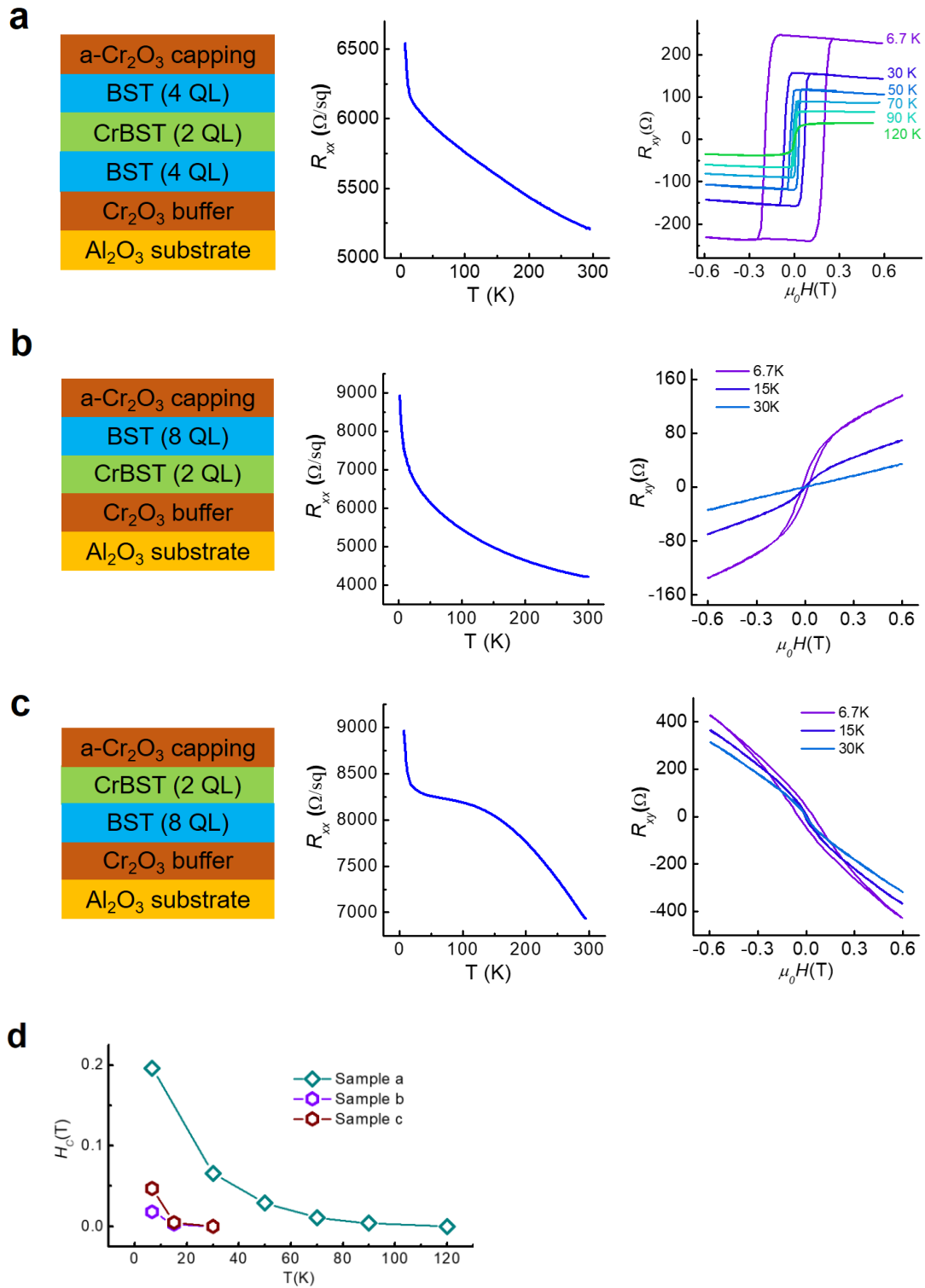

Figure S5: (a-c) Schematic structure, temperature-dependent longitudinal resistance and field-dependent Hall resistance of three samples with the same composition but different stacking sequence. (d) Coercive fields determined from (a-c).

S6

## 6. Comparison of Hall effect for t = 0 sample over time

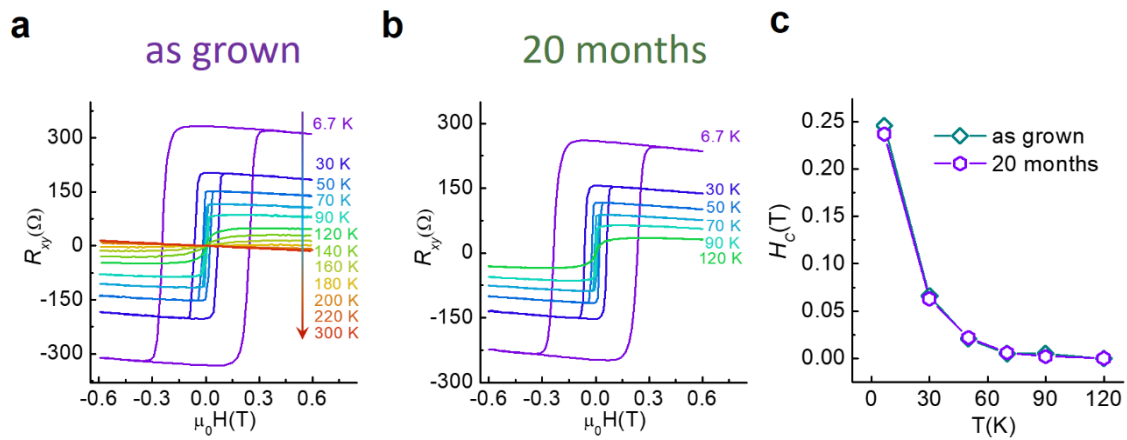

Figure S6: Field-dependent Hall resistance measured at different temperatures for sample t = 0 (a) as grown and (b) 20 months later. (c) Coercive fields determined from (a) and (b).

# 7. Control samples for testing the relationship between THE and $Cr_2O_3$ buffer/capping layer

**a**
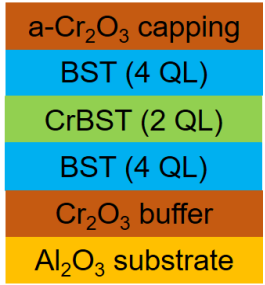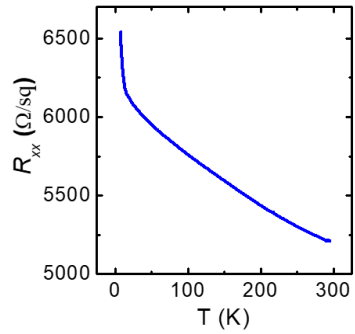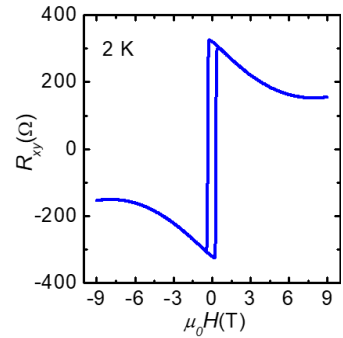

**b**
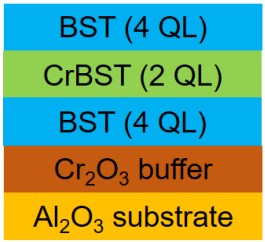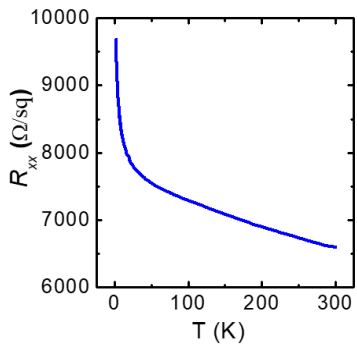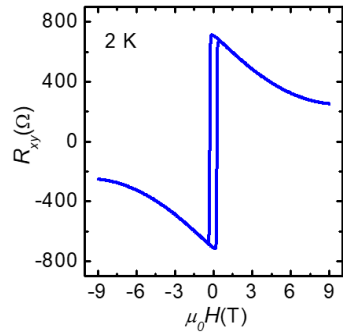

**c**
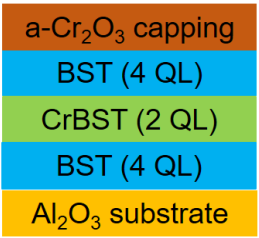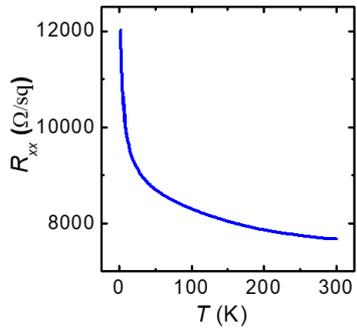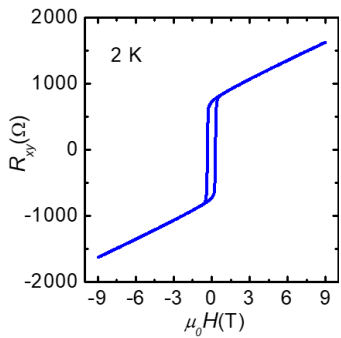

**d**
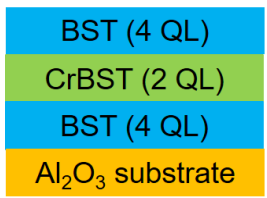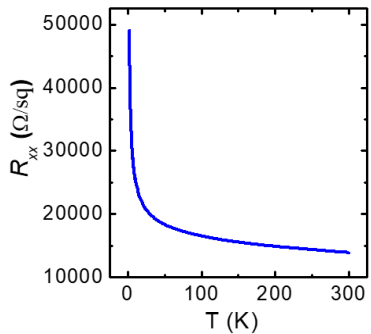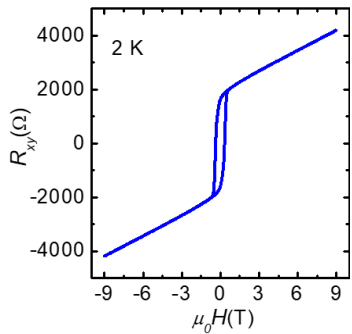

Figure S7: Schematic structure, temperature-dependent longitudinal resistance and field-dependent Hall resistance of a series of control samples with or without $Cr_2O_3$ buffer or capping layer.

## 8. Fitting of the topological Hall effect

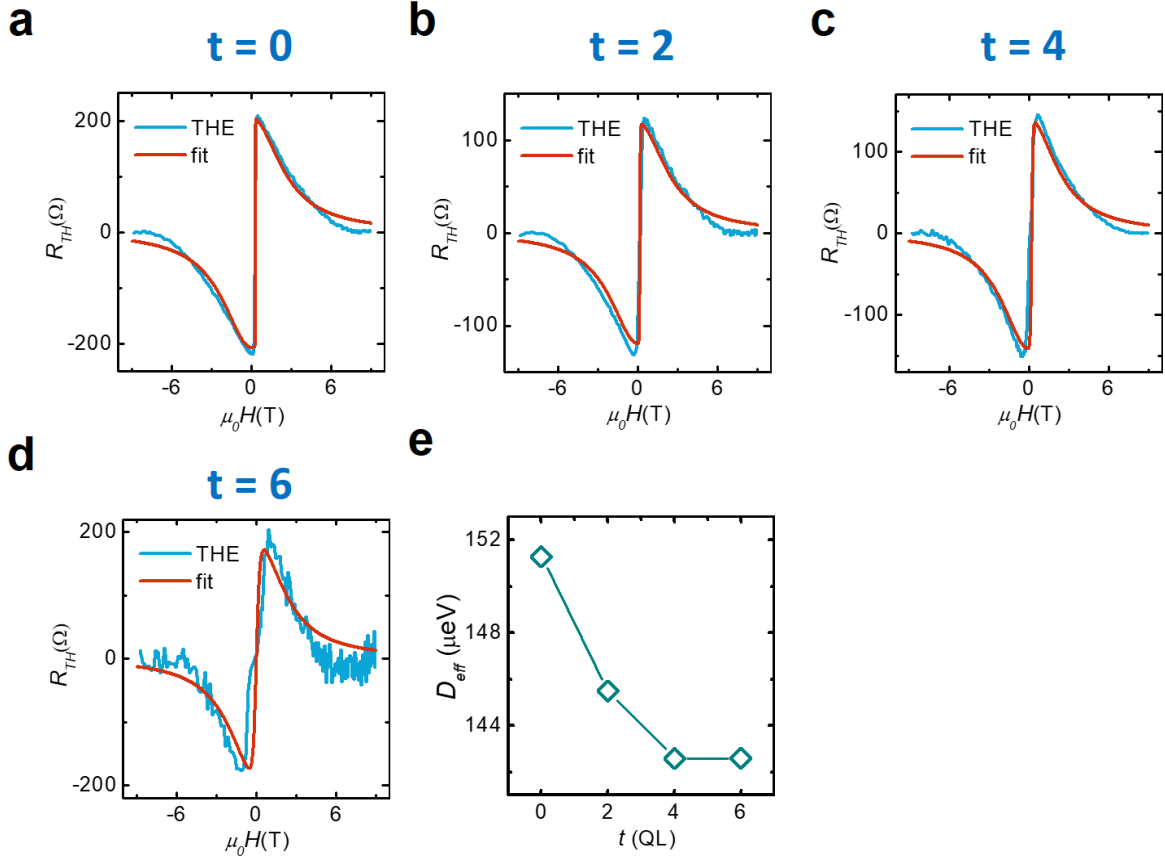

Figure S8: Fitting of topological Hall resistance to $\frac{M(H)}{1+(H/D_{eff})^2}$. Considering that M(H) data is unavailable, we used the anomalous Hall fitting in Figure 4 to approximate M(H). The fitted $D_{eff}$ values for t = 0, 2, 4, and 6 are 2.61 T, 2.51 T, 2.46 T and 2.46 T, which corresponds to energy scales of 151 μeV, 146 μeV, 143 μeV and 143 μeV, respectively, as shown in (e).

## 9. Hall resistance data of sample t = 8 before and after subtracting ordinary Hall resistance

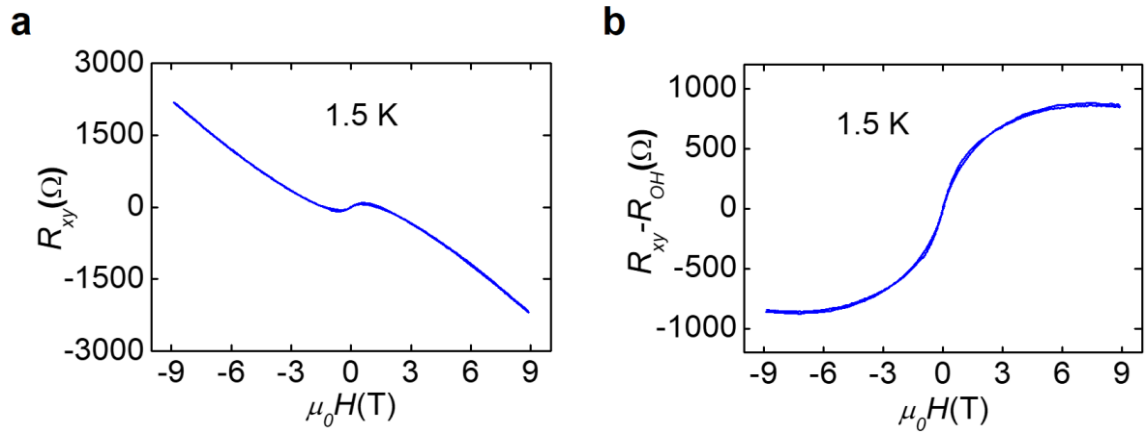

Figure S9: Hall resistance of sample t = 8 measured at 1.5 K before (a) and after (b) subtracting ordinary Hall resistance.

## 10. Transport data of a homogeneously Cr-doped BST sample

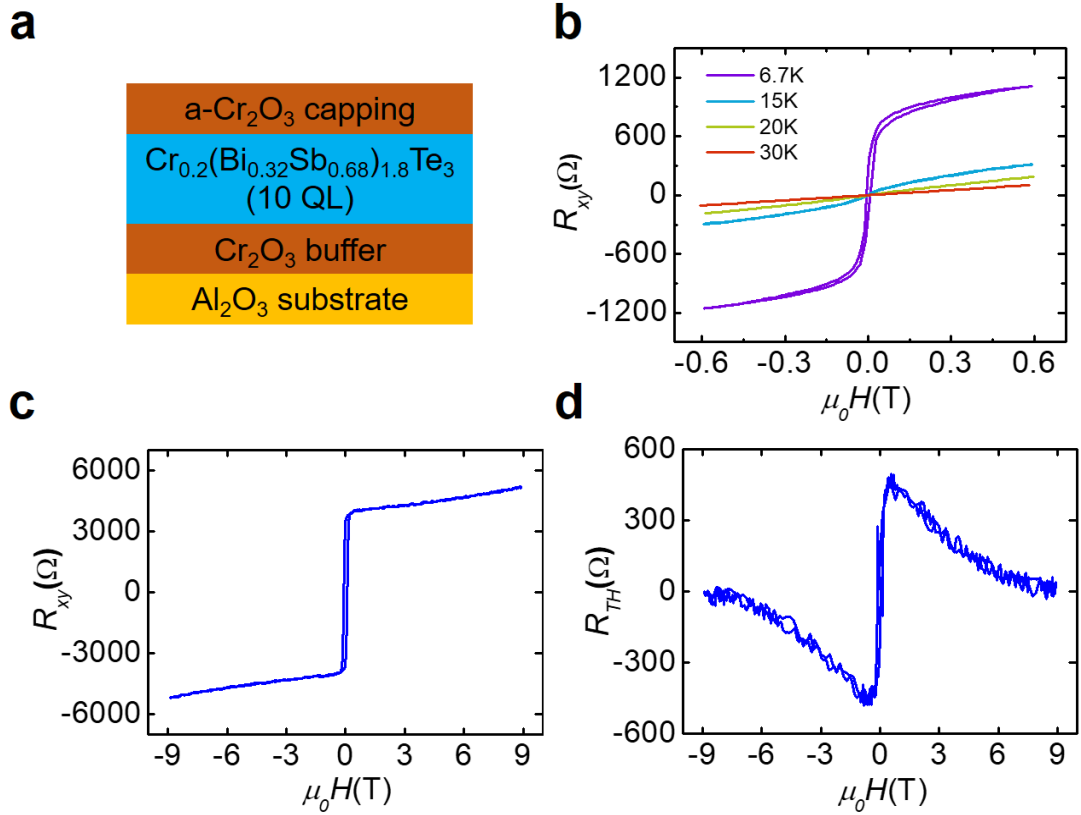

Figure S10: (a) Schematic of the sample structure. (b) Hall resistance measured at different temperatures under magnetic fields up to 0.6 T. (c) Hall resistance measured at 1.5 K and magnetic field up to 9 T. (d) Topological Hall resistance after subtracting ordinary Hall resistance and anomalous Hall resistance in (c).